\documentclass{PoS}
\usepackage{subcaption}
\def\beq{\begin{equation}}
\def\eeq{\end{equation}}
\usepackage{tikz}
\usepackage{amsbsy}
\title{Exotic Phases of a Higgs-Yukawa Model with Reduced Staggered Fermions}
\ShortTitle{Higgs-Yukawa Model}

\author{\speaker{Simon Catterall}$^a$, Nouman Butt$^a$ and David Schaich$^b$ \\
\llap{$^a$}Department of Physics, Syracuse University, Syracuse, New York 13244, United States\\
\llap{$^b$}Department of Mathematical Sciences, University of Liverpool, \\ Liverpool L69 7ZL, United Kingdom

E-mail: \email{smcatter@syr.edu}, \email{ntbutt@syr.edu}, \email{david.schaich@liverpool.ac.uk}
}

\abstract{
We investigate the phase structure of a four dimensional SO(4) invariant lattice Higgs-Yukawa model comprising four reduced staggered fermions interacting with a real scalar field. The fermions belong to the fundamental representation of the symmetry group while the three scalar field components transform in the self-dual representation of SO(4). We explore the phase diagram and find evidence of a continuous transition
between a phase where the fermions
are massless to one where the fermions acquire mass.  This transition
is not associated with symmetry breaking and there is no obvious local order parameter.}

\FullConference{37th International Symposium on Lattice Field Theory - Lattice2019\\
		16-22 June 2019\\
		Wuhan, China}

\begin{document}
\section{Lattice Action and Symmetries}
Conventionally fermions acquire a mass gap through either explicit or spontaneous
symmetry breaking. It is well known that certain Higgs-Yukawa
lattice models can be constructed that gap fermions without breaking
symmetries using strong four fermion interactions. However until
recently these symmetric massive phases have been regarded as
lattice artifacts separated from the continuum by broken symmetry phases and/or first order phase transitions.
The model described in this proceedings seems to circumvent these constraints and appears to offer the possibility of
allowing for a continuum limit in which fermions acquire mass without breaking symmetries. Apart from the
intrinsic interest in such a dynamics it opens up the possibility of formulating chiral lattice fermions using
the Eichten-Preskill approach described in \cite{Eichten:1985ft}.

The action for the model is
\beq
S = \sum_{x} \psi^a [ \eta.\Delta^{ab} + G\sigma^{+}_{ab} ] \psi^{b} + \frac{1}{4}\sum_{x} (\sigma^{+}_{ab})^2 - \frac{\kappa}{4} \sum_{x,\mu} \left[ \sigma^{+}_{ab}(x) \sigma^{+}_{ab}(x + \mu) + \sigma^{+}_{ab}(x) \sigma^{+}_{ab}(x-\mu) \right]
\eeq
where repeated indices are to be contracted and $\eta^{\mu}(x)=\left(-1\right)^{\sum_{i=1}^{\mu-1}x_i}$ are the usual staggered fermion phases.
The discrete derivative is
\begin{equation}
\Delta_{\mu}^{ab} \psi^{b} = \frac{1}{2}\delta^{ab} [\psi^{b}(x +\mu) - \psi^{b}(x-\mu)].
\end{equation}
The self-dual scalar field $\sigma^{+}_{ab}$ is defined as
\begin{equation}
\sigma^{+}_{ab} = P^{+}_{abcd} \sigma_{cd} = \frac{1}{2} \left[ \sigma_{ab} + \frac{1}{2} \epsilon_{abcd} \sigma_{cd}\right]
\end{equation}
with $P^{+}$ projecting the antisymmetric matrix field $\sigma(x)$ to its self-dual component which transforms under only
one of the SU(2) factors comprising SO(4). It is the presence of the additional SU(2) symmetry of the fermion
operator that allows one to prove the associated Pfaffian operator is real, positive semi-definite and hence the
model can be simulated without encountering a sign problem.

When $\kappa=0$ we can integrate out the scalar field and obtain the pure four fermi model
studied in Refs.~\cite{Ayyar:2016lxq, Catterall:2016dzf}. While a single peak is observed in a certain fermion
susceptibility the eventual consensus of this earlier work was
that this peak straddled a very narrow symmetry broken phase bordered
by two phase transitions~\cite{Schaich:2017czc}.  In Ref.~\cite{Catterall:2017ogi} it was argued that it should
be possible to eliminate this broken phase by tuning an additional scalar kinetic term. In concrete terms it should
be clear from the form of the action that $\kappa>0$
favors ferromagnetic ordering of the scalar field and associated fermion bilinear. This is to be contrasted
with the preferred antiferromagnetic ordering observed in the earlier pure four fermi work. Hence by tuning $\kappa$ it should be logically possible to reach a phase with no spontaneous ordering of the fermion bilinear. Furthermore, general RG arguments
suggest that the renormalizability of the theory should necessitate the addition of such a term. Results for this model
appeared in Ref.~\cite{Butt:2018nkn}.\footnote{In 3d RG arguments do not require such a term and indeed no intermediate phase is seen~\cite{Ayyar:2015lrd}.}

In addition to its manifest SO(4) symmetry, the action is also invariant under a shift symmetry
\begin{equation}
\psi(x) \to \xi_{\rho} \psi(x + \rho)
\end{equation}
with $\xi_\mu(x)=\left(-1\right)^{\sum_{i=\mu}^dx_i}$
as well as a discrete $Z_2$ symmetry:
\begin{eqnarray}
\sigma^{+}& \to& - \sigma^{+} \\
\psi^a &\to& i\epsilon(x) \psi^a.
\label{eq:Z_2}
\end{eqnarray}
Both of the $Z_2$ and SO(4) symmetries prohibit local bilinear fermion mass terms from appearing as a result of quantum corrections.
While non-local SO(4)-symmetric bilinear terms can be constructed by coupling fields at different sites in the unit hypercube, such terms break the shift symmetry. Further detailed discussion of possible bilinear mass terms is presented in Ref.~\cite{Catterall:2016dzf}.

\section{Numerical Results}
\begin{figure}
  \begin{subfigure}[b]{0.49\textwidth}
    \includegraphics[width=\textwidth]{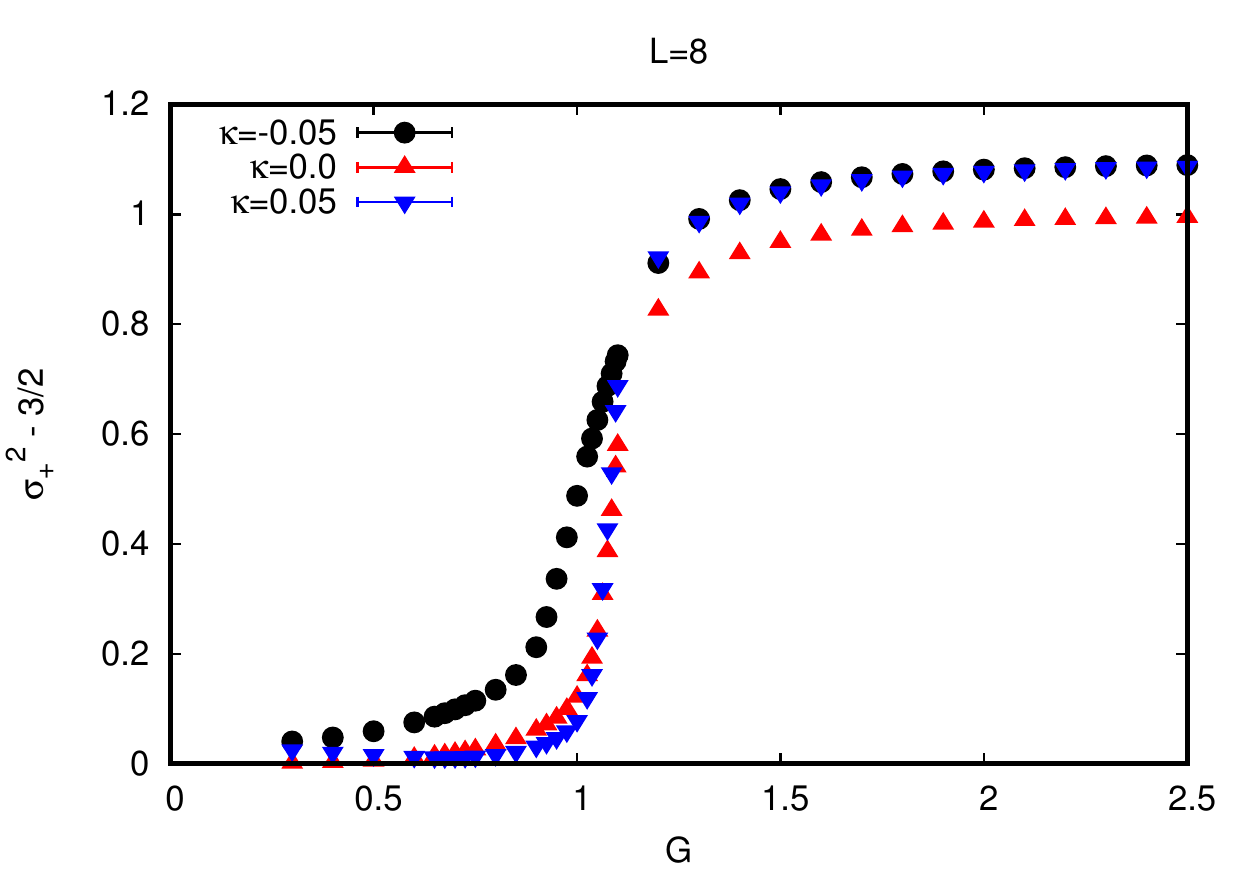}
    \caption{Four fermion condensate vs $G$}
    \label{plus}
  \end{subfigure}
  \hfill
  \begin{subfigure}[b]{0.49\textwidth}
    \includegraphics[width=\textwidth]{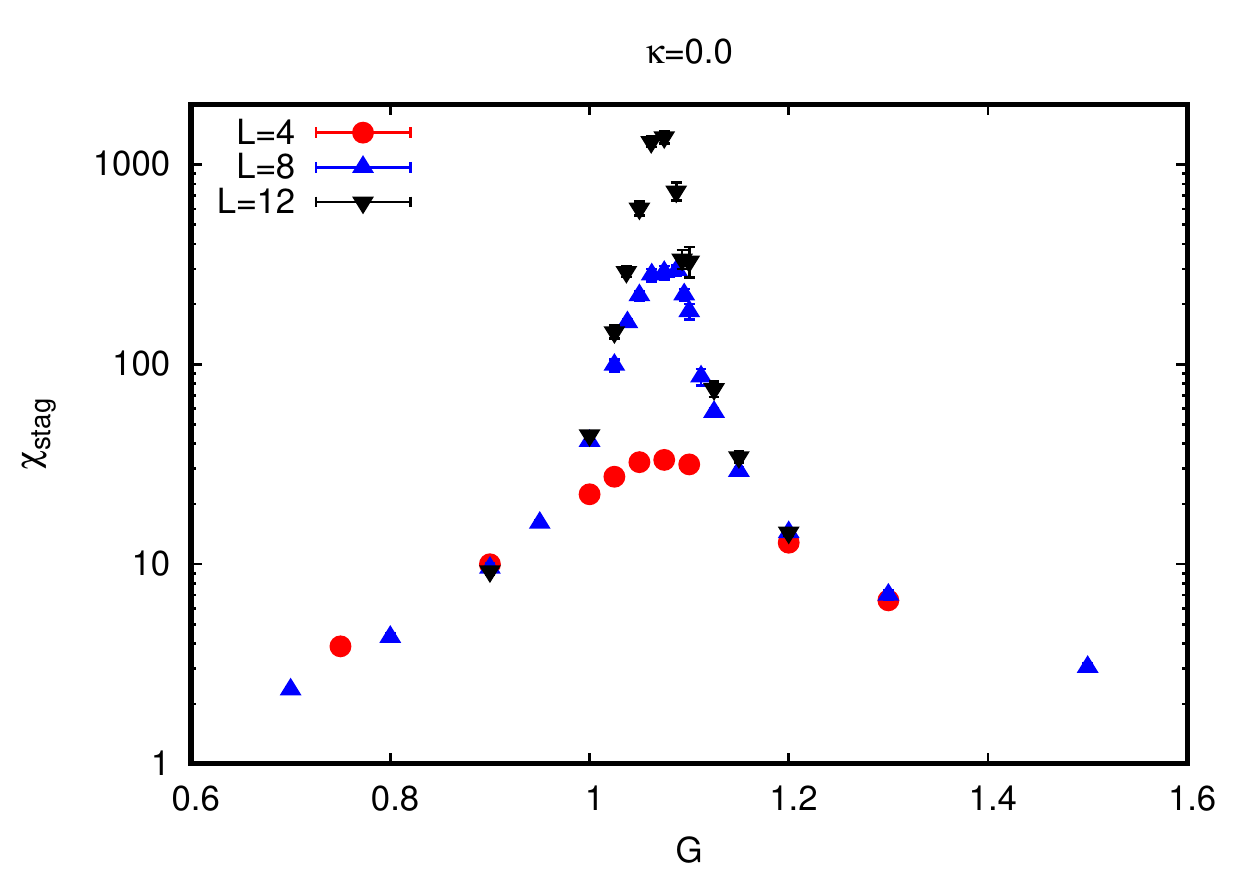}
    \caption{Susceptibility at $\kappa=0$}
    \label{susk0}
  \end{subfigure}
\end{figure}

Despite these constraints fermions can nevertheless pick up a mass via
interaction with a {\it symmetric} four fermion condensate.  The latter is shown in
fig.~\ref{plus} which plots the vev of the SO(4) symmetric operator
$O=\epsilon^{abcd}\psi_a\psi_b\psi_c\psi_d$
for three values of $\kappa=-0.05,0.0-0.05$ as $G$ is varied.
A strong coupling expansion of the $\kappa=0$
theory results in a prediction for the momentum space fermion propagator of the form
\beq
  F(p)=\frac{\sqrt{6G^2} i\sin{p_\mu}}{\sum_\mu \sin^2{p_\mu}+m_F^2}
\eeq
where a non-zero fermion mass $m_F^2=24G^2-8$ is generated.

However, a fermion mass can also arise via a spontaneous breaking of one of
the exact symmetries and the formation of a bilinear condensate.
This appears to happen at $\kappa=0$ where an intermediate antiferromagnetically ordered
phase appears around $G\sim 1.05$ - see fig.~\ref{susk0} which shows a staggered fermion susceptibility that scales
linearly with volume as expected in the presence of a condensate. This susceptibility is defined by
\beq
  \chi=\frac{1}{V}\sum_{x,y}\epsilon(x)\epsilon(y) \psi^a(x)\psi^b(x)\psi^a(y)\psi^b(y).
\eeq
To study this question more carefully we have
added an SO(4) breaking source term to the action and studied the condensate as a function of the magnitude of
the source as the thermodynamic limit is taken.
The source terms take the form
\begin{equation}
  \label{eq:d}
  \delta S = \sum_{x,a,b} (m_1 + m_2 \epsilon(x) ) [ \psi^a(x)\psi^{b}(x) ] \Sigma^{ab}_{+}
\end{equation}
where the SO(4) symmetry breaking source $\Sigma^{ab}_{+}$ is
\begin{equation}
  \Sigma^{ab}_{+} = \left(\begin{array}{cc}i\sigma_2 & 0\\
                                           0         & i\sigma_2\end{array}\right).
\end{equation}

\begin{figure}
  \begin{subfigure}[b]{0.49\textwidth}
    \includegraphics[width=\textwidth]{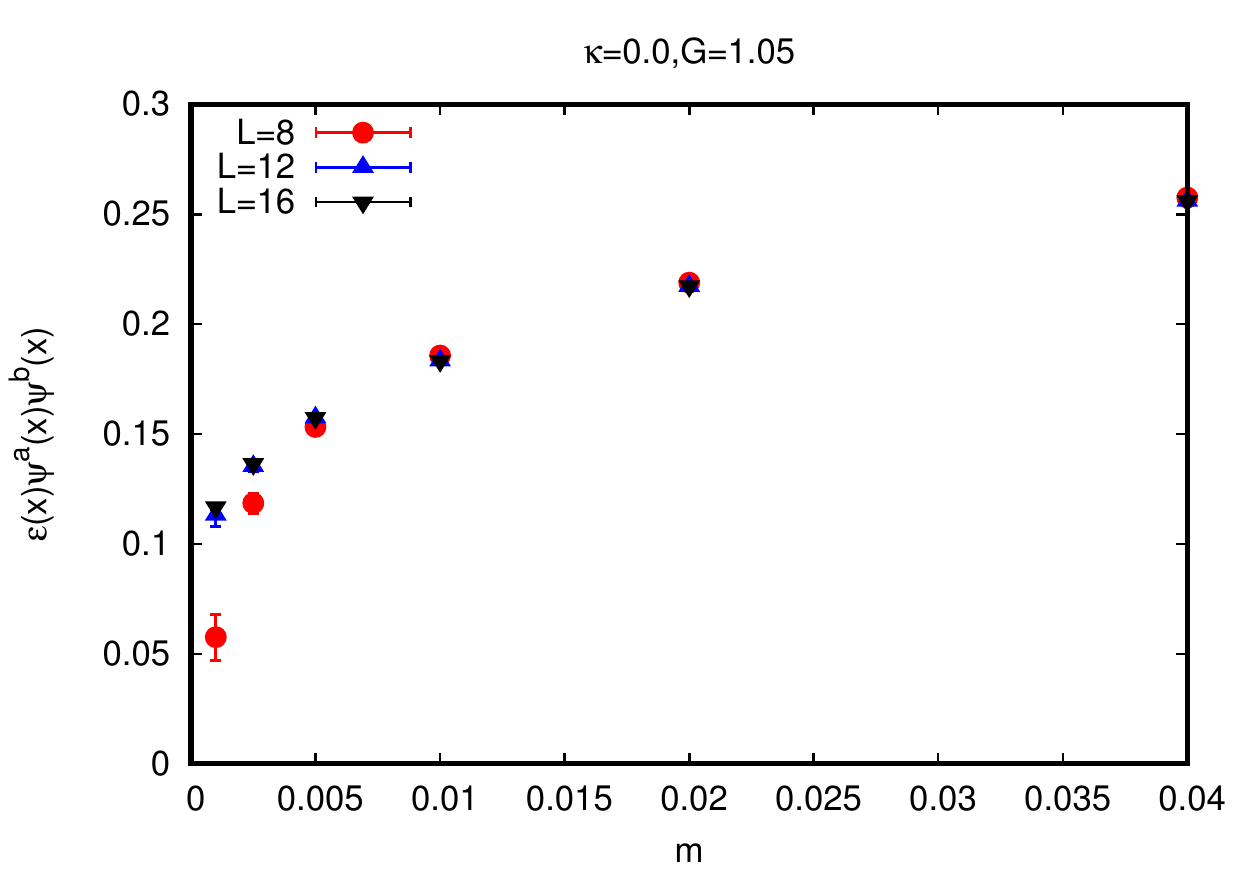}
    \caption{Condensate vs $m_2$ at $\kappa=0$}
    \label{bilink0}
  \end{subfigure}
  \hfill
  \begin{subfigure}[b]{0.49\textwidth}
    \includegraphics[width=\textwidth]{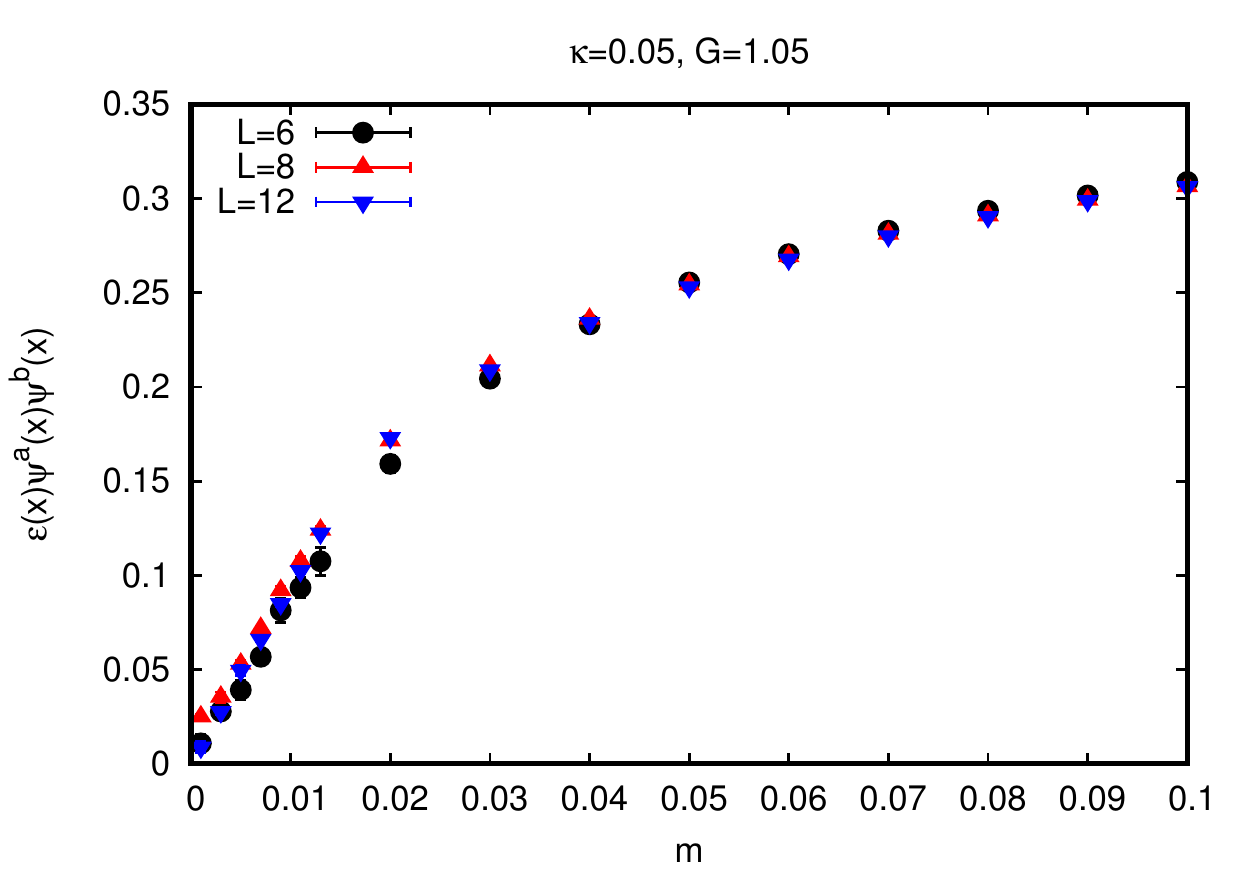}
    \caption{Condensate vs $m_2$ at $\kappa=0.05$}
    \label{bilink005}
  \end{subfigure}
\end{figure}

The results for the vev of the antiferromagnetic bilinear ($m_1=0$)
as a function of source are shown in fig.~\ref{bilink0} at $\kappa=0$ and
$G=1.05$ corresponding to the peak of the susceptibility. Clearly for small values of the source the signal grows with increasing volume. This effect trades off
against the general result that the vev must vanish for zero source on a finite volume. As a result a plateau develops
that strengthens and moves to smaller source values as the thermodynamic limit is taken. This provides unequivocal evidence for
spontaneous symmetry breaking at this value of $G$ and confirms the presence of a condensate at this point in the
phase diagram.

\begin{figure}
\begin{center}
\includegraphics[width=0.75\textwidth]{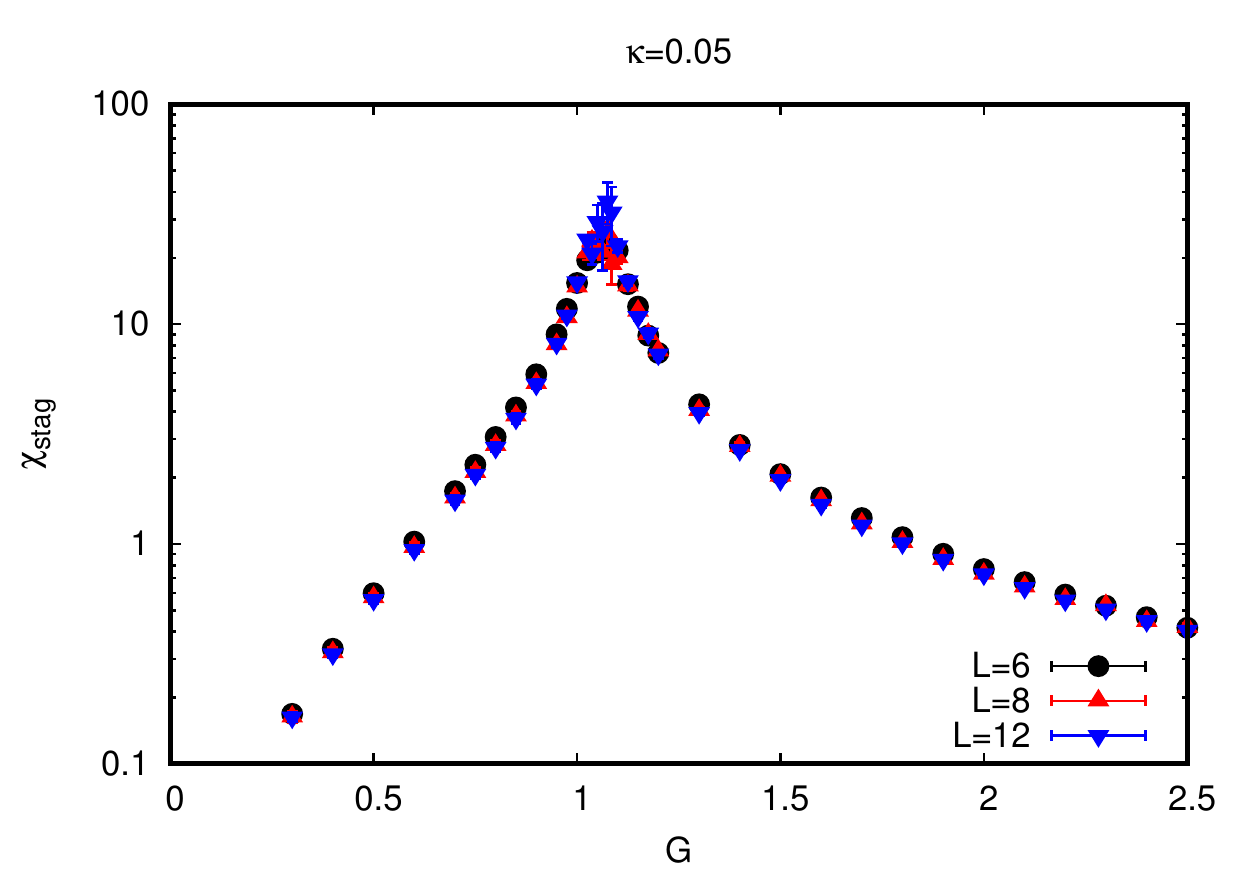}
\caption{Susceptibility at $\kappa=0.05$\label{susk005}}
\end{center}
\end{figure}

\begin{figure}
\begin{center}
\includegraphics[width=0.75\textwidth]{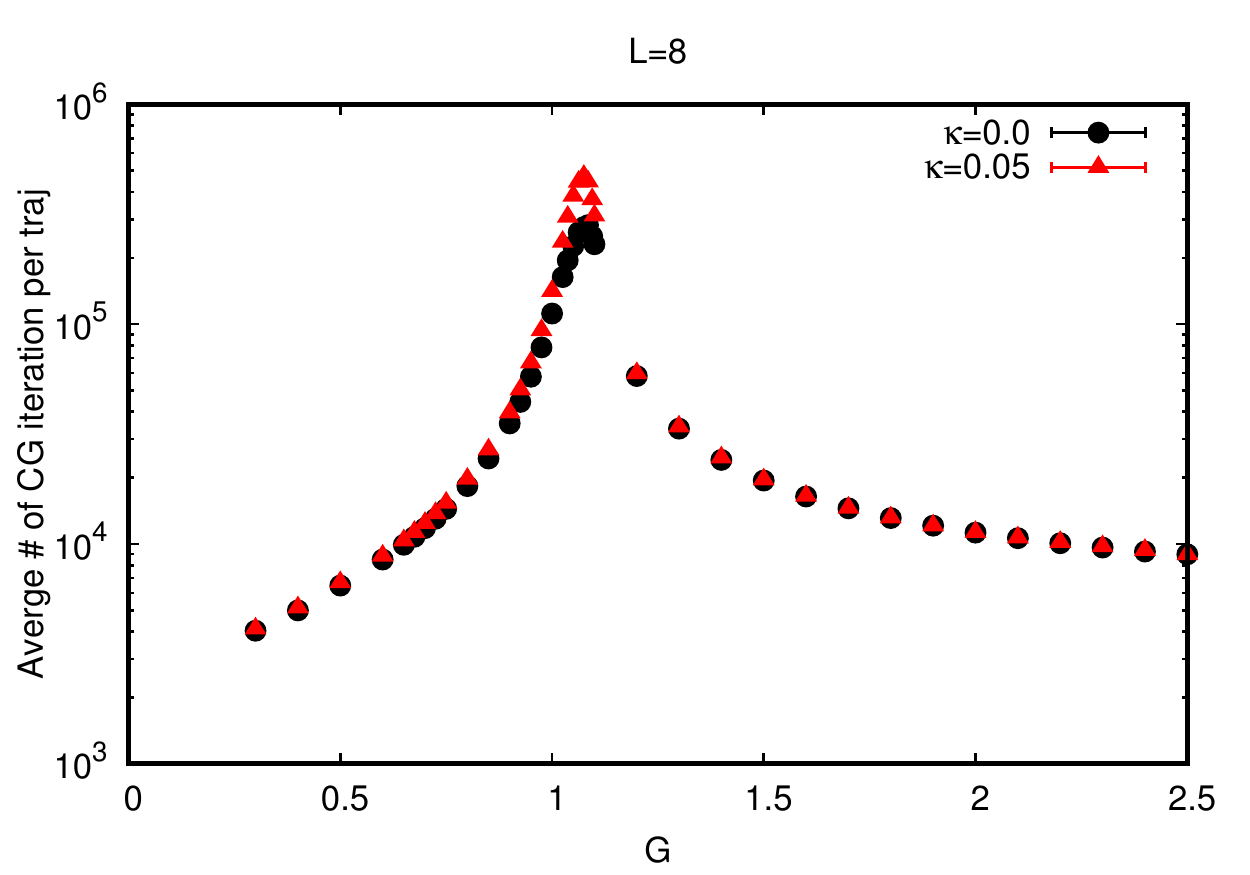}
\caption{Number of CG iterations vs $G$ for $\kappa=0$ and $\kappa=0.05$ at $L=8$\label{cg}}
\end{center}
\end{figure}

Contrast this with fig.~\ref{bilink005} which shows the behavior of the same quantity at $\kappa=0.05$
and $G=1.05$.
In the latter plot the bilinear shows no significant volume dependence and falls smoothly to zero as the source is reduced.
Thus in this region of the phase diagram there is no evidence of symmetry breaking. Of course one should worry that
perhaps there is no phase transition at all or that the transition
is shifted away from $G\sim1.05$.  Fig.~\ref{susk005} shows the corresponding
susceptibility which reveals a peak at that same value of the Yukawa coupling $G$. However, notice that
this peak is independent of the volume. While this is consistent with the absence of a
condensate it leaves open the possibility that only a crossover rather than a true phase transition survives for $\kappa>0$. One way to test this is
to examine
the number of conjugate gradient iterations required
to invert the fermion operator close to the transition. Fig.~\ref{cg} shows a striking peak close to $G\sim 1.05$
similar to that seen at $\kappa=0$  and confirms the system still appears to undergo a phase
transition for $(G,\kappa)=(1.05,0.05)$. Indeed the peak is {\it higher} for positive $\kappa$ indicating a faster growth
of the correlation length as compared to the $\kappa=0$ situation. Thus the numerical evidence favors
a continuous phase transition at small, positive $\kappa$ and no intermediate, broken phase.

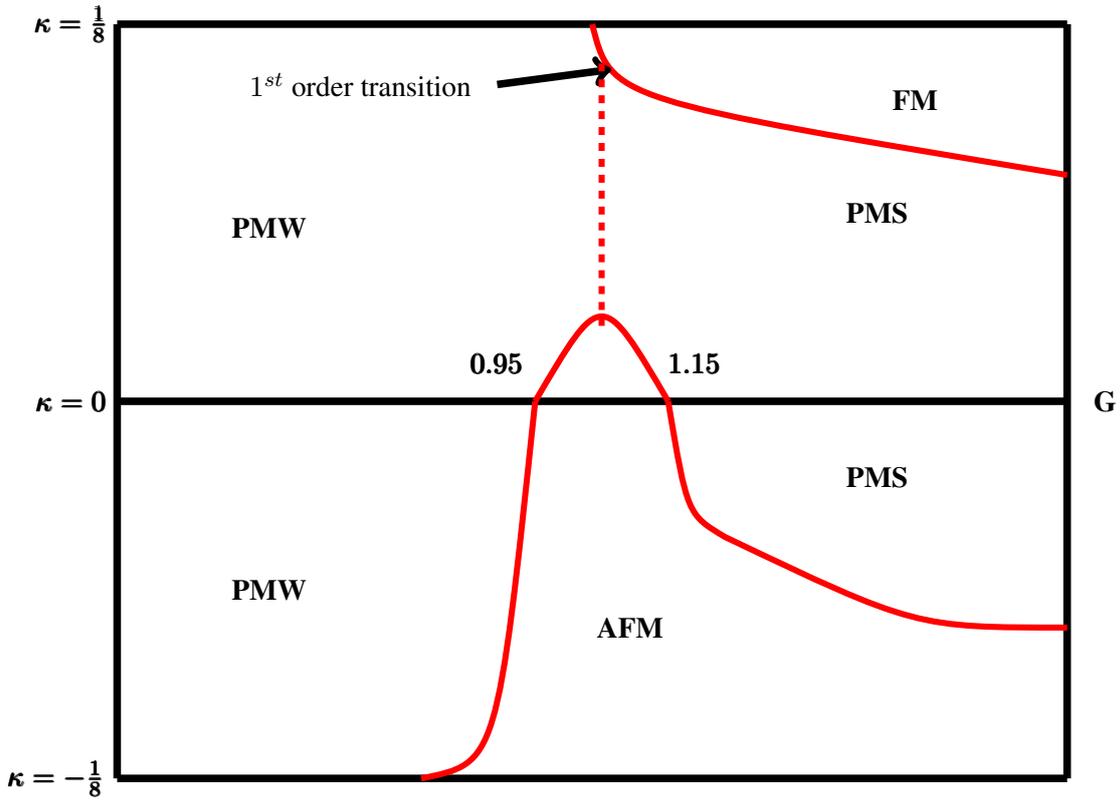
\begin{figure}
  \begin{minipage}[c]{\textwidth}
    \begin{tikzpicture}
    \draw [line width=1.0mm,black] (0,-5) -- (0,0) -- (0,5) ;
    \draw [line width=1.0mm,black] (0,0) -- (12.5,0) ;
    \draw [line width=1.0mm,black] (0,5) -- (12.5,5) ;
    \draw [line width=1.0mm,black] (0,-5) -- (12.5,-5) ;
     \draw [line width=1.0mm,black] (12.5,-5) -- (12.5,5) ;
     \draw[red,line width =0.8mm ] (6.25, 5) .. controls (6.5,4.0) .. (12.5,3);
     \draw[red,line width =0.8mm ] (7.25, 0) .. controls (7.5,-1.5) .. (8.0,-1.8) .. controls (10.5,-3.0) .. (12.5,-3);
     \draw[red,line width =0.8mm ] (5.5, 0) .. controls (5.0,-4.8) .. (4.0,-5);
     \draw[red,line width =0.8mm ] (5.5, 0) .. controls (6.375,1.5) .. (7.25,0);
     \draw [->,black, line width=1.0mm] (5.0,4.2) -- (6.5,4.4);
     \draw [red,line width=0.8mm ,dashed] (6.375,1.0) -- (6.375,4.5);
     \draw node[scale=1.0] at (3.2,4.2) {$1^{st}$ order transition};
     \draw node[scale=1.0] at (2.0,2.3) {\textbf{PMW}};
      \draw node[scale=1.0] at (2.0,-2.5) {\textbf{PMW}};
       \draw node[scale=1.0] at (10.0,2.5) {\textbf{PMS}};
        \draw node[scale=1.0] at (10.0,-1.0) {\textbf{PMS}};
        \draw node[scale=1.0] at (5.0,0.5)  {\pmb{$0.95$}};
        \draw node[scale=1.0] at (7.6,0.5)  {\pmb{$1.15$}};
        \draw node[scale=1.0] at (6.75,-3.0)  {\textbf{AFM}};
         \draw node[scale=1.0] at (10.5,4.0)  {\textbf{FM}};
         \draw node[scale=1.0] at (13.0,0) {\textbf{G}};
         \draw node[scale=1.0] at (-0.6,5.0)  {\pmb{$\kappa=\frac{1}{8}$}};
         \draw node[scale=1.0] at (-0.6,0.0)  {\pmb{$\kappa=0$}};
         \draw node[scale=1.0] at (-0.8,-5.0)  {\pmb{$\kappa=-\frac{1}{8}$}};
    \end{tikzpicture}
  \end{minipage}
  \caption{Sketch of the phase diagram in the $(\kappa,G)$ plane\label{fig:phase}}
\end{figure}

We have explored this model over a wider range of couplings and find results consistent with the phase diagram
sketched in fig.~\ref{fig:phase}. For $\kappa<0$ the antiferromagnetic (AFM) phase observed for $\kappa=0$ widens and still
separates a massless symmetric phase (the so-called paramagnetic weak or PMW phase) from a massive
symmetric phase (the paramagnetic strong coupling or PMS phase). If $\kappa$ is sufficiently large
and positive we see signs of a ferromagnetic (FM) phase, which for large $G$ is separated from the PMS phase by
first order phase transitions. The most interesting region is the one we focused on above, namely
small positive $\kappa$ and Yukawa couplings close to unity. In this region there appears to be a continuous transition
between the two symmetric phases. It is important to recognize that some sort of phase transition must be present at this point since on one
side the fermion is massless while on the other it is gapped.

\section{Summary}
Our numerical results, which were obtained with a Rational Hybrid Monte Carlo algorithm, appear to reveal the
existence of a continuous phase transition separating a massless symmetric (PMW) from a massive symmetric
(PMS) phase in the plane of Yukawa and kinetic couplings. We have concentrated the bulk of our efforts on the region near
$(G,\kappa)=(1.05,0.05)$.
In the original four fermi model with $\kappa = 0$, this value of $G = 1.05$ lies close to the center of the antiferromagnetic phase.
The transition between massless and massive phases appears to be continuous around $(G,\kappa)=(1.05,0.05)$ and there is no sign of a bilinear fermion condensate either
antiferromagnetic or ferromagnetic in nature at that point.  If the evidence for this transition holds up on larger
lattices it suggests the existence of a new fixed point for strongly coupled fermions in four dimensions and should
allow for the possibility of giving fermions mass in the continuum without breaking symmetries. This is
already a very interesting possibility.

It also may offer some hope for formulating chiral fermions on the lattice following an old idea by
Eichten and Preskill in which strong four fermion interactions are used to gap out would be doubler states
that arise in a naive approach to formulating Weyl fermions on the lattice. The basic  idea is to modify the
Yukawa terms by including additional derivative operators so that the interactions become large for
doubler states but remain small for the primary, naive Weyl fermions. The spirit of the approach is rather similar to the inclusion of
a Wilson term for vector like theories in which lattice derivatives are inserted into a bare fermion mass term
to generate large doubler masses. For a chiral theory there are no invariant mass terms but there are
invariant four fermion or Yukawa couplings and provided these break any anomalous
symmetries it is possible to imagine a scenario where the doublers can
be decoupled.  It would be very interesting to repeat the work described here for such a model
based on reduced staggered fermions to see whether such a scenario can in fact be realized.


\section*{Acknowledgments}
SMC acknowledges the support of the DOE through the grant DE-SC0009998.
DS was supported by UK Research and Innovation Future Leader Fellowship MR/S015418/1.

\end{document}